# Free Standing Epitaxial Oxides Through Remote Epitaxy: The Role of the Evolving Graphene Microstructure

Asraful Haque*, Suman Kumar Mandal, Shubham Kumar Parate, Harshal Jason D'souza, Sakshi Chandola, Pavan Nukala, Srinivasan Raghavan*.

Center for Nanoscience and Engineering, Indian Institute of Science, Bengaluru.

**Abstract:**

Remote epitaxy has garnered considerable attention as a promising method that facilitates the growth of thin films that replicate the crystallographic characteristics of a substrate by utilizing two-dimensional (2D) material interlayers like graphene. The resulting film can be exfoliated to form a freestanding membrane and the substrate, if expensive, can be reused. However, atomically thin 2-D materials are susceptible to damage before and during film growth in the chamber, leading to a poor epitaxy. Oxide remote epitaxy using graphene, the most commonly available 2D material, is particularly challenging because the conventional conditions employed for the growth of epitaxial oxides also degrade graphene. In this study, we show for the first time that a direct correlation exists between the microstructure of graphene, it's getting defective on exposure to the pulsed laser deposition plume, and the crystalline quality of the barium titanate film deposited on top. A controlled aperture method was used to reduce graphene damage. Even so, the degree of damage is more at the graphene grain boundaries than within the grains. Large grain-sized >300 microns, graphene suffered less damage and yielded a film comparable to that grown directly on a strontium titanate substrate with a rocking curve half width of 0.6°. Using large grain-sized bi-layer graphene, 4 mm x 5 mm oxide layers were successfully exfoliated and transferred onto SiOx-Si. These insights pave the way for the heterogeneous integration of functional oxides on foreign substrates, holding significant implications for commercializing perovskite oxides by integrating them with Si-CMOS and flexible electronics.

**Introduction:**

Oxides form a class of materials exhibiting diverse properties, including high dielectric constant, piezo and ferroelectricity, and ferromagnetism [1]. The growth of crystalline oxides on silicon and other substrates is crucial for seamlessly integrating various functional devices like sensors, memories, and MEMS onto the cost-effective Si CMOS platform[2]. However, this is challenging due to a native amorphous oxide layer on silicon, which impedes epitaxial growth. To overcome this challenge, McKee et al. first introduced a sub-monolayer of alkaline earth metal (e.g., Sr or Ba), which deoxidizes the native oxide upon heating at elevated temperature and thereafter allows deposition of an epitaxial $SrTiO_3$ layer [3]. However, this method needs sophisticated ultra-high vacuum tools such as molecular beam epitaxy (MBE). Other similar methods have included using thermodynamically more stable oxides than $SiO_x$, such as MgO [4] and $ZrO_2$ [5] or nitrides (TiN) [6], as buffer layers before depositing the required functional oxide.

In addition to integration with Si, the surging demand for wearable devices necessitates the incorporation of crystalline oxides onto flexible substrates [7]. The crystallization temperature of these oxides surpasses the melting point of flexible substrates, rendering the utilization of oxide thin films in flexible devices problematic. The layer transfer method resolves these problems [8,9]. This involves transferring the functional oxide from the growth substrate to another substrate of interest. This substrate could be a single crystal oxide substrate that does not have issues related to growth on Si and also yields films with good structural quality. However, on such single-crystal oxide substrates, the traditional method of direct growth and detachment cannot be used for layer transfer due to the strong bonding of the film with the substrate. Other approaches like sacrificial layers, mechanical spalling, interface melting, and substrate etching have material choice, cost, interface quality, and throughput limitations [10–13]. Such single-crystal oxide substrates are also expensive and need to be reused.

A possible solution to both requirements, oxide substrates, and reusability, involves the use of freestanding single crystalline oxide membranes obtained by employing 2D materials, such as graphene, at the interface during growth [8]. The film nucleates on the 2D material but orients itself via the interatomic potential emanating from the substrate and permeating through it, leading to the so-called remote epitaxy mode of growth [14]. However, the mechanism underlying remote epitaxy remains contentious due to the potential for lateral overgrowth through pinholes or defects in the 2D material [15]. Distinguishing between these mechanisms is a challenging and actively pursued endeavour. Nevertheless, using 2D materials undeniably offers a proven method for obtaining single-crystalline freestanding membranes, boasting advantages such as minimal material waste (no need for sacrificial layers), expensive substrate reusability, and integration onto flexible and other desired substrates [16–18].

Achieving successful remote epitaxy of films necessitates careful consideration of factors including (a) the choice of 2D material, (b) selection of substrate material, (c) interaction between the film and the substrate at a certain thickness of the 2D material, (d) the method for producing the 2D material-coated single-crystalline substrate, either via direct growth or transfer and (e) extend of damage to the 2D material during thin film growth itself. Graphene is the prevalent choice for a 2D material due to its well-established large-scale growth capability to achieve monolayer thickness [19,20]. Substrate choice is crucial, as it determines the extent of electrostatic potential penetration. Functional oxides with ionic bonds exhibit the most robust ionic potential penetration compared to III-As and III-N compounds [21]. This holds dual utility: (a) enabling the use of thicker 2D materials (i.e., multilayer graphene at the interface) and (b) facilitating the employment of wet transfer methods for graphene where the interface is not as clean as dry transferred graphene [22]. However, dry transfer results in cleaner interfaces and better ionic potential transparency; the high ionicity of the oxide substrate ensures that even wet-transfer graphene will be effective [22]. Unlike dry transfer methods that require high-temperature graphitization furnaces and costly SiC substrates, a wet transfer is feasible using chemical vapor deposited (CVD) polycrystalline graphene grown easily on Cu foils [23,24].

Pulsed laser deposition (PLD) is the most common method for obtaining epitaxial oxide films. However, preserving the integrity of the graphene layer during heat up and PLD growth of oxide thin films, where the graphene is exposed to elevated temperatures, oxygen-rich environments, and plasma plumes, remains challenging. This study aims to understand the role of the evolving graphene microstructure on the quality of freestanding epitaxial oxide films obtained by remote epitaxy using PLD. In the process of doing so, we accomplished, for the first time, PLD remote epitaxy of BaTiO$_3$ (BTO) on STO substrate using wet transferred graphene and its successful exfoliation. Previous reports on remote epitaxy of BTO use either MBE [21] (which is much more complicated) or PLD, which either uses the dry transfer of graphene [23] (where costly SiC substrates are used to graphitize the top layer at a much higher thermal budget of >2000 °C) or as grown graphene on the single crystalline substrate [25] (which is not universal for all substrates) (Figure S1 shows a comprehensive summary from the literature). We observed that the defect density in graphene after exposure to the BTO PLD plume depends on its as-grown microstructure. As has been suggested previously [26], to minimize damage to the graphene layer during the rigorous growth conditions of BTO, we used a two-step growth of BTO, with and without a slit. Despite doing so, we see that large, >300 µm, grain-sized graphene suffers less damage than small, < 10 µm, grain-sized graphene. For large grain sizes (>300 µm), it is further observed that while there is damage to graphene in general, it is more concentrated at the grain boundaries. This, in turn, affects the crystalline quality of BTO by a factor of 1.5X, as measured by x-ray rocking curves. Two graphene layers are required for successful exfoliation of the epitaxial oxide layer. Strain relaxation and crystallography of the exfoliated films have been studied in depth. The exfoliated films maintain their crystalline order and are ferroelectric, as depicted by PFM measurements. Unlike previous reports of oxide remote epitaxy utilizing dry-transfer graphene [23,27], our approach yields epitaxial BaTiO$_3$ membrane via wet-transferred graphene for possibly the first time, to the best of our knowledge (see Figure S2(a-f) for wet-transfer process details).

**Result and Discussion**

**Damage during PLD and the role of graphene microstructure:**

The epitaxial PLD growth conditions for ferroelectric BTO on STO lie in the temperature range of 700-800 °C and $10^{-3}$ to $10^{-2}$ mbar O$_2$ pressure regime [13,28]. Such high temperature and oxygen pressures are expected to severely affect the structural integrity of the graphene layer and, hence, the remote epitaxy of BTO. Therefore, investigation of the microstructural evolution that the graphene layer undergoes in such conditions is necessary. The $I_D/I_G$ ratio, representing the intensity ratio of the D and G peaks in Raman spectroscopy, and the Full Width at Half Maxima (FWHM) of the D peak will be used to monitor defect levels in graphene [16,29]. A lower $I_D/I_G$ (<0.5) [24,30] ratio indicates fewer defects and higher quality graphene. To study the stability of graphene at 700 °C and ~5x$10^{-6}$ mbar, typical ramp-up conditions pre-growth in our PLD chamber, a graphene/SiO$_2$ substrate (wet transferred graphene on SiO$_2$ with distinct G and 2D Raman peaks, with an $I_{2D}/I_G$ ~ 2, confirming the presence of monolayer graphene without any observable D peak, indicating minimal defects, Figure S2) was loaded into the PLD chamber with a base pressure of 1x$10^{-6}$ mbar. The choice of SiO$_2$ as a substrate was deliberate, as it exhibits reduced fluorescence compared to crystalline substrates, thereby enabling precise and unambiguous measurements of graphene's Raman signals. Subsequently, the chamber pressure was elevated to $5\times10^{-6}$ mbar, and the substrate temperature was raised to 700°C. The graphene/SiO$_2$ substrate was maintained at this temperature for 25 minutes, referred to as the pre-treatment process. After that, it was taken out of the PLD chamber for examination. Raman spectroscopy was conducted at room temperature, reaffirming the presence of graphene from its G and 2D peaks without any observable D peak (refer to Figure 1(b)), thereby indicating that the ramp-up to growth conditions has no deleterious effect on graphene microstructure. The morphology of the graphene layer was investigated using AFM. Figure 1(c, d) shows a 1µm x 1µm AFM image of a graphene/SiO$_2$ before and after pre-treatment. Wrinkles and PMMA residues (PMMA being used during the transfer process) can be observed in the sample before pre-treatment, whereas after pre-treatment, no PMMA residues are present, but only folds/wrinkles can still be seen. The RMS roughness reduces from 1.25 nm to 0.26 nm post pre-treatment. To avoid PMMA use, we also tried dry transfer of graphene from Cu foil onto SiO$_2$ using a PDMS stamp (see Figure S3); however, graphene cracking during stamping was observed. Therefore, wet-transferred graphene was chosen for further sample fabrication since graphene remains protected with minimal residue post-pre-treatment at 700°C and $10^{-6}$ mbar pressure.

To evaluate the extent of damage to the graphene layer upon BTO deposition (i.e., due to PLD plume), additional graphene-coated SiO$_2$ samples were prepared and subjected to PLD deposition of BTO (using a fluence of 1.04 J/cm$^2$ and growth rate of 0.08 nm/pulse) under different ambient conditions of $P_{N2}=5\times10^{-3}$ mbar, $P_{Ar}=5\times10^{-3}$ mbar, and at a pressure of $5\times10^{-6}$ mbar (without purging any gas) at 700°C. O$_2$ background gas was intentionally avoided to eliminate the chances of graphene oxidation. Subsequent Raman spectroscopy measurements conducted at room temperature and presented in Figure 1(b) exhibited the presence of G and 2D peaks in all samples. The 2D and G band is the signature of the presence of graphitic sp$^2$ materials [31]. However, a notable D (defect) peak, attributed to damage from the plasma plume, emerged in graphene with a full width at half maximum (FWHM) between 61.3 ± 1.5 cm$^{-1}$ to 89.7 ± 1.3 cm$^{-1}$ (see Figure S4 and Table TS1) and $I_D/I_G$ ratio of 1.2±0.04 to 1.6±0.07.

Therefore, a two-step deposition method was adopted to mitigate the damage to the graphene layer during BTO deposition (as depicted in the accompanying Figure 1(a)). In this approach, the initial BTO layer was grown at $5\times10^{-6}$ mbar pressure and 700°C, with a minimal growth rate of 0.02 nm/pulse and laser fluence of 1.04 J/cm$^2$. Subsequently, the thicker BTO film was grown with higher growth rates of 0.08 nm/pulse, an $O_2$ pressure of 5 x $10^{-3}$ mbar, a temperature of 750 °C, and a constant laser fluence of 1.04 J/cm². A rectangular slit at the laser source was employed to achieve a laser spot area of 5.2 mm² during the first layer, while a laser spot area of 17.5 mm² was used for the second one. The FWHM of the defect peak in graphene post-deposition is 24.2 ± 0.6 cm$^{-1}$, (see Figure S4, Table S1) compared to 68.3 ± 1.1 cm$^{-1}$ (without a slit, Table S1) and $I_D/I_G$ ratio of 1.3±0.1. Studies by Lee et al. showed that reducing the laser spot size decreases the density of ionic species and their kinetic energy [26,32]. This reduction in ionic density and its kinetics results in a significant decrease in ionic bombardment onto graphene and hence a lower defect peak FWHM of 24.2 ± 0.6 cm$^{-1}$ compared to 68.3 ± 1.1 cm$^{-1}$ (without a slit, Table S1). Thus, in contrast to a one-step growth method, a two-step growth technique utilizing a slit-based approach proves advantageous in minimizing damage to the graphene layer. This is in agreement with another recent report by M. A. Wohlgemuth et al.[26].

CVD graphene can be synthesized with a wide variety of grain sizes and grain boundary structures. It is very well known that grain boundaries are the weak points and are susceptible to damage [24,33]. The work described so far was done using graphene with a grain size of 6 microns. To study the impact of grain size, graphene layers with sizes in excess of 300 microns were prepared. Here, readers are referred to Figure S5 under section II of supplementary information for details of the growth steps of large and small grain size full coverage graphene on copper foil (also Figure S6 for the method to determine grain size). It is worth emphasizing that both the small and the large grain-sized graphene were annealed in methane post monolayer graphene formation to ensure that graphene grain boundaries are as defect-free as possible [24]. Thus, it is expected that the only difference between the small- and large-grained graphene is the density of grain boundary and triple junction-related regions per unit area and not the defect structure of the boundaries themselves. BTO was grown on the large grain size graphene (~322 µm) coated $SiO_2$ substrate.

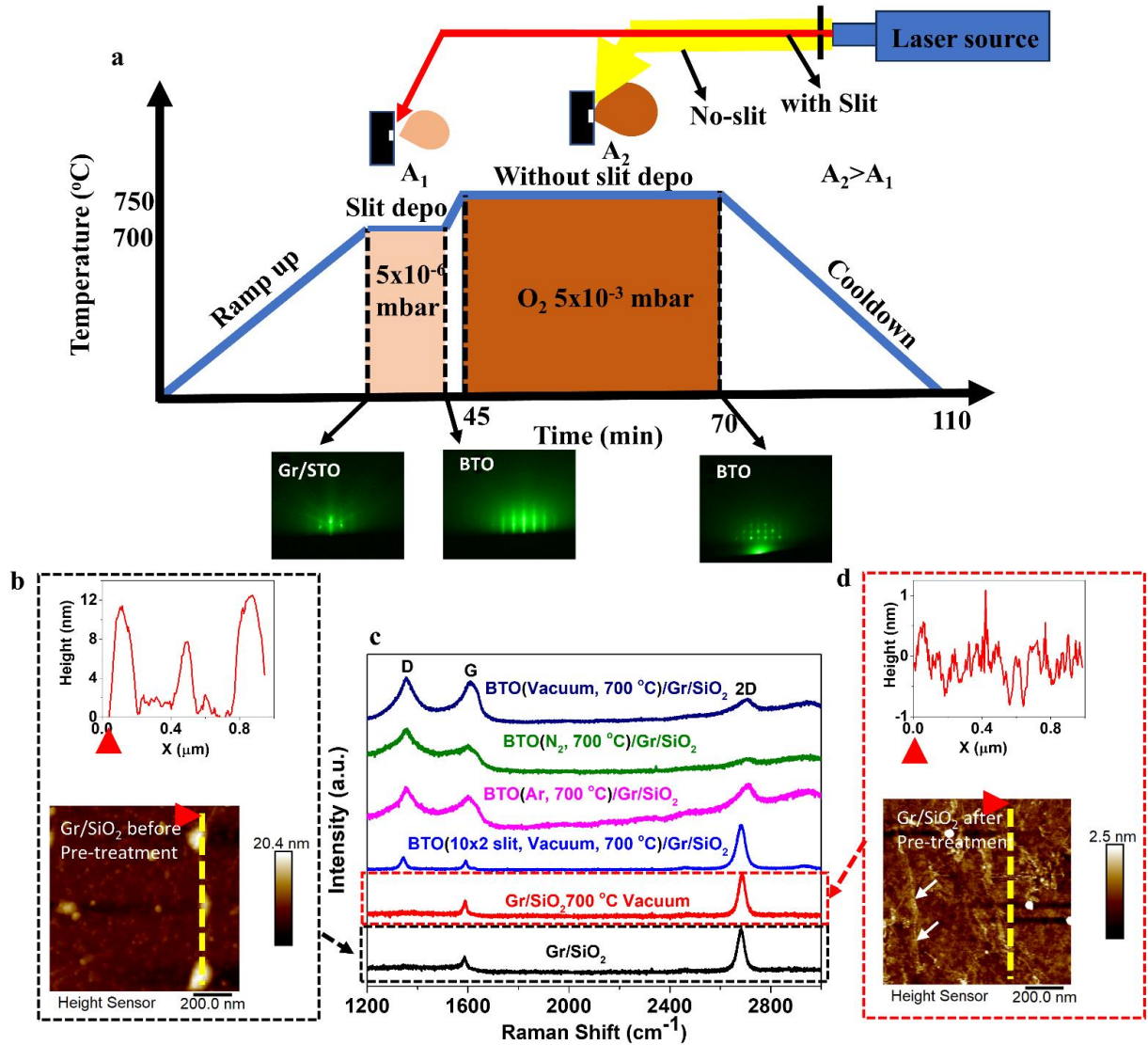

**Figure 1: Schematic of two-step slit-based BTO deposition process and effect of PLD parameters on graphene on SiO$_2$.** (a) Schematic representation of the two-step slit-based BTO deposition process. The lower panel showcases Reflective High Energy Electron Diffraction (RHEED) patterns at distinct stages. (b) AFM micrograph captured before pre-treatment (pre-treatment at a temperature of 700 °C and a pressure of 5x10$^{-6}$ mbar), revealing the residual presence of PMMA after the wet transfer of graphene. (c) Raman spectrum of graphene on SiO$_2$ (progressing from bottom to top): before pre-treatment, after pre-treatment, post-BTO deposition (at 700 °C, pressure 5x10$^{-6}$ mbar) with the utilization of a slit at the laser source, post-BTO deposition (at 700 °C in an Argon (Ar) ambient pressure of 5x10$^{-3}$ mbar), post-BTO deposition (at 700 °C in a Nitrogen (N$_2$) pressure of 5x10$^{-3}$ mbar), and post-BTO deposition (at 700 °C, and pressure of 5x10$^{-6}$ mbar) without the implementation of a slit at the laser source. Independent samples were employed for each iteration. The Full Width at Half

Maximum (FWHM) values of defect peaks are tabulated in Table TS1 of the supplementary information. (d) AFM micrograph after pre-treatment, demonstrating the absence of PMMA residues following the wet transfer of graphene. Notably, wrinkles and folds are discernible (indicated by white arrows).

Notably, the transferred graphene onto $SiO_2$ exhibits a similar $I_D/I_G$ ratio regardless of the graphene's grain size (Figure 2(a)). However, following exposure to the BTO plume, the $I_D/I_G$ ratio of small-grained (1.3±0.1) graphene surpasses that of its large-grained (0.8±0.1) counterpart (Figure 2(a)). Figure 2(b) shows the $I_D/I_G$ line scan along the BTO/large graphene//$SiO_2$ sample showing peaks in the $I_D/I_G$ at intervals of typical grain size of large-grained graphene, which confirms the grain boundaries of graphene are defective sites, rendering them more susceptible to damage during exposure to the PLD plume. This phenomenon has also been validated by a study conducted by J Y Lee et al.[33], where they imaged graphene grain boundaries by selectively etching them in oxygen plasma followed by thermal oxidation of the underlying copper substrate along graphene grain boundaries.

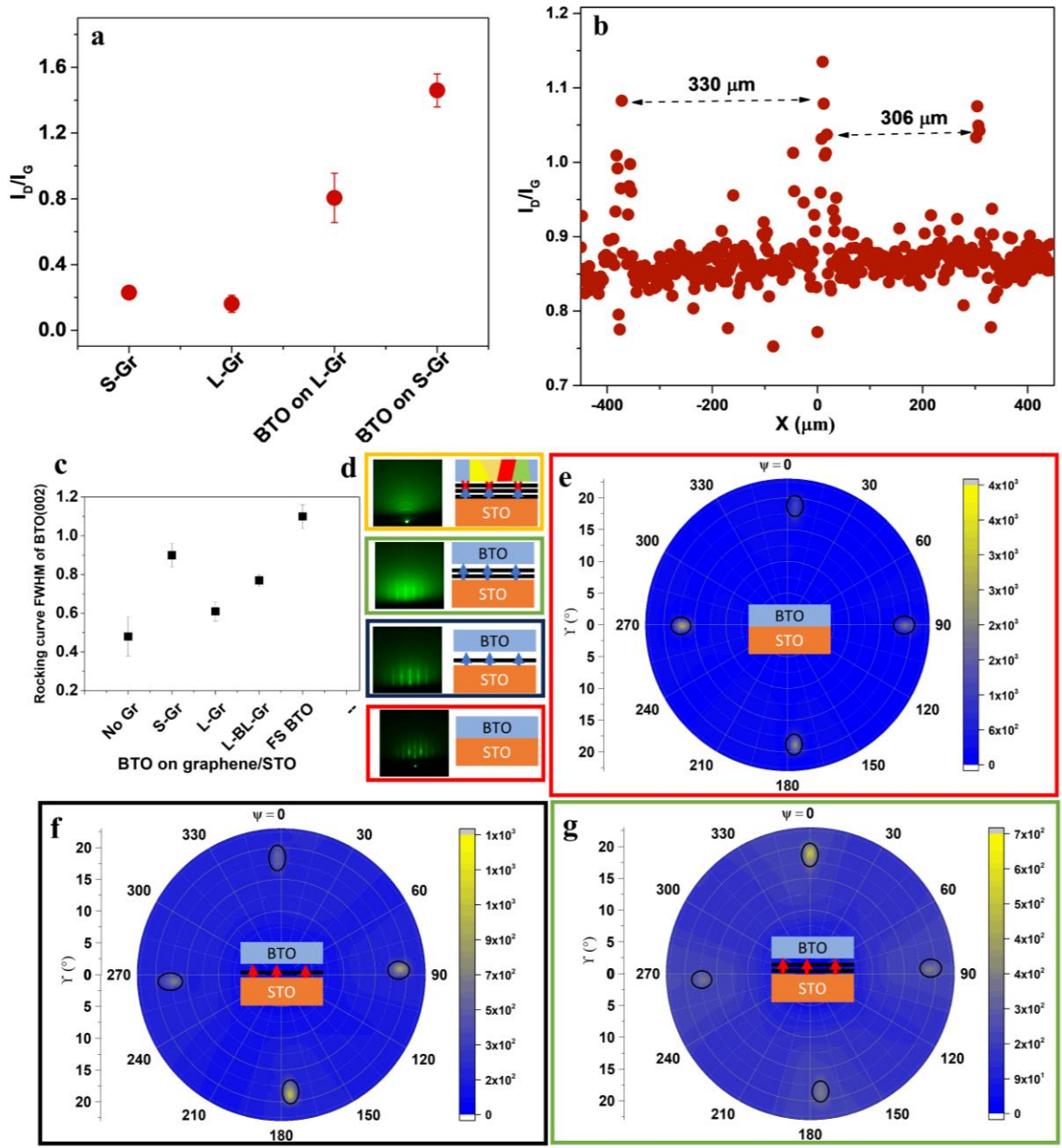

**Figure 2:** (a) $I_D/I_G$ ration obtained from Raman Spectrocopy of small-grained graphene //(SiO$_2$, large-grained graphene//SiO$_2$, BTO/small-grained graphene//SiO$_2$, BTO on large-grained graphene//SiO$_2$ (average taken from several points) (b) $I_D/I_G$ line scan for 900 μm length on BTO/large-grpahene//SiO$_2$ sample. (c) XRD Rocking curve FWHM of BTO (002) grown on: STO, small grain graphene (s-gr)~ 6 μm//STO, large grain graphene (L-gr)~ 322μm //STO, bilayer large grained graphene (L-BL-gr)//STO and freestanding (FS) BTO. (d) RHEED pattern and schematic of BTO grown on STO, on monolayer large-grained graphene//STO, on bilayer large-grained graphene//STO, and trilayer large-grained graphene// STO. Pole figure (PF) around the (103) peak of BTO film grown on (e) STO, (f) (L-gr)//STO, (g) (L-BL

gr)//STO. The radial direction represents χ, which ranges from 0-19°, while the azimuthal direction represents phi, with a (0-360°) range. Color represents intensity (right of each PF).

**Structural characterization of BTO concerning layer of graphene and its microstructure:**

Graphene-coated STO substrates (obtained by wet transfer of CVD graphene from Cu foil to STO substrate) were prepared for epitaxial integration of BTO. The RHEED pattern of the graphene//STO substrate, depicted in Figure S7 (a, b), exhibits a luminous yet diffused specular reflection in contrast to the bare STO RHEED pattern under similar conditions. This diffused specular reflection originates from the randomly oriented uppermost graphene layer. Feeble diffraction streaks (indicated by an arrow) are ascribed to the underlying STO substrate (visible in Figure S7 (b)). Following this, BTO (~100 nm thickness) was grown by PLD on both STO and the graphene//STO substrate. A high-resolution x-ray diffraction (HRXRD) θ-2θ scan across the 10-80° range presents diffraction peaks of the (00l) plane series (Figure S7(c)), indicating that the film is oriented in the out-of-plane direction. In-plane orientation is addressed later. Notably, the rocking curve full-width half-maximum (RC FWHM) of BTO (002) Bragg reflection increases from 0.48° ±0.04° in the BTO//STO to 0.90°±0.06° in the BTO/6-micron grain sized graphene//STO as shown in Figure 2(c) (Figure S8). This observation stands in contrast to the findings of Yuwei Guo et al. [34], wherein remote epitaxially PLD-grown $VO_2$/graphene//sapphire structures exhibit a smaller symmetric RC FWHM of 0.47° in comparison to 0.63° for directly grown $VO_2$//sapphire references [34]. In their case, this reduction in FWHM was attributed to the role of the graphene layer in mitigating epitaxial strain within the film, thus resulting in diminished defect formation, such as dislocations, and improved crystalline quality [34].

Now, we discuss the role of the graphene microstructure itself. Repeating the above deposition by using the >300 microns grain-sized graphene yielded a reduction in the symmetric RC FWHM of (002) BTO reflection from 0.9°±0.06° in BTO/6 micron grain-sized graphene//STO to 0.61°±0.05° in BTO/>300 micron grain sized graphene//STO Figure 2(c). However, the RC FWHM of 0.61°±0.05° (in the BTO/large grain graphene//STO configuration) still exceeds the value of 0.48°±0.04° (in the BTO//STO case), which could be attributed to defect formations like folds and wrinkles during the wet transfer of graphene (as observed in Figure 1(d)).

As seen in Figure S1, the literature points towards using multiple layers of wet or dry-transferred graphene layers for successful exfoliation of the functional oxides. Having successfully deposited epitaxial BTO on STO utilizing a monolayer of graphene, it is crucial to understand the effect of multiple graphene layers on the epitaxial quality of BTO grown on top and its correlation with the facilitation of its exfoliation process. Therefore, to elucidate the evolution of strain relaxation with the number of graphene layers (monolayer ML, bilayer BL, and trilayer TL), graphene-coated STO substrates were prepared by repeatedly transferring large grain graphene (average grain size 322 µm), followed by BTO growth employing a slit-based two-step deposition approach as previously discussed. The crystallographic characteristics of BTO/(ML, BL, TL) large graphene//STO substrates were examined via X-ray diffraction. Figure S9(a) shows the specular out-of-plane (OP) XRD $\theta$-$2\theta$ scan of BTO/(ML, BL, TL) large graphene//STO and BTO//STO in the proximity of STO (002) Bragg reflection. The appearance of BTO (002) diffraction peaks adjacent to the STO (002) reflection (at $2\theta = 46.4°$) is evident in the samples with ML and BL graphene at the interface, as well as in the directly grown BTO on STO. Further confirmation of crystalline quality and orientation is derived from the RHEED pattern of the BTO surface, confirming crystallinity and orientation of BTO in BTO//STO, BTO/ML- large graphene//STO, BTO/BL-large graphene//STO; however, the TL- large graphene coated STO sample manifests polycrystalline characteristics of BTO (as observed in Figure 2d). This is attributed to the weak potential field of STO beyond the second layer, leading to a loss of substrate registry. The assessment of remote epitaxial BTO film crystallinity is depicted through the rocking curve in Figure 2(c). The FWHM of the rocking curve shifts from 0.48°±0.04° for BTO//STO to 0.61°±0.05° for BTO/ML large grain graphene//STO to 0.77°±0.03° for BTO/BL large grain graphene//STO as shown in (Figure 2(c), Figure S8).

These results, however, raise the question of whether BTO is following the atomic registry of the graphene layer (van der Waal epitaxy) or that of the substrate (remote epitaxy). Figure 3(e-g) addresses this question through XRD pole figure (PF) analyses using the (103) Bragg diffraction condition, showcasing four spots separated by $\varphi = 90°$ and at $\chi = 18.4°$ for the BTO grown directly on STO and with ML and BL graphene at the interface. Consequently, the epitaxial relationship of [001]BTO//[001]STO and (100)BTO// (100)STO is established in all three cases, negating the possibility of van der Waals epitaxy.

Graphene stability using two-step slit-based PLD deposition was established earlier in Figure 1(a). Further, energy-dispersive X-ray spectroscopy (EDX) analysis conducted on BTO/BL Gr//STO verifies a distinct interface. Figure 3(a) shows that the graphene layer's carbon (C) element signal can be identified at the BTO//STO interface. Additionally, Raman mapping over a 6 μm x 6 μm area indicates the distribution of D, G, and 2D peaks, affirming the presence of graphene after BTO growth, Figure 3(b, c).

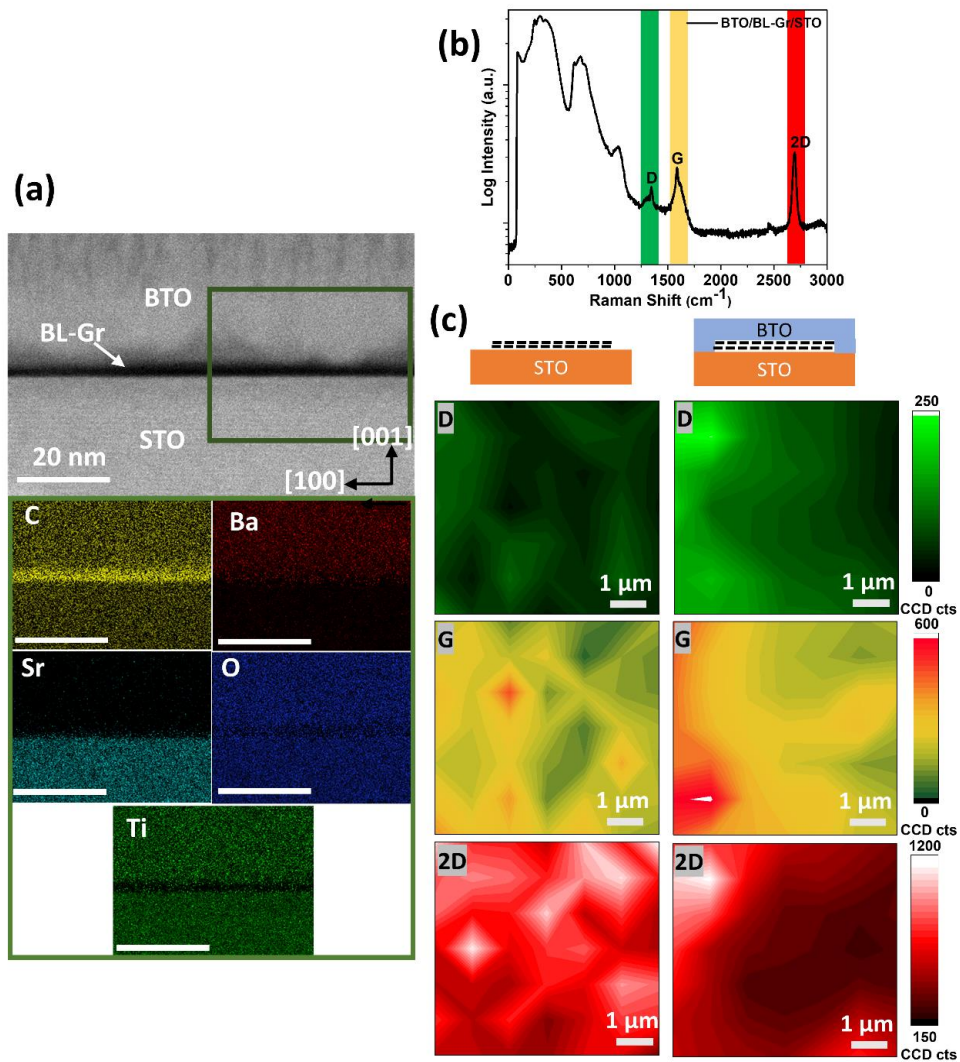

**Figure 3** (a) cross-sectional HAADF-STEM image and EDS mapping of BTO grown on bilayer graphene-coated STO substrate (b) Raman spectra between 0-3000 cm$^{-1}$ of BTO/BL Gr//STO substrate indicating graphene G and 2D peaks, (c) Raman mapping showing the intensity of D, G, and 2D peak of graphene scanned in a region of 6 μm x 6 μm area for BL-graphene on STO and BTO/BL-graphene//STO.

**Flexible membrane and its transfer:**

Considering the forces present at the interface of BTO/graphene//STO, both weaker interactions through the graphene and stronger ones where the graphene is defective or absent due to tearing, exfoliation, and transfer of the remote epitaxial BTO film were undertaken. Figure 4(a-d) outlines the comprehensive process, with the first stage involving the deposition of a Ni stressor layer (refer to methods for details), followed by exfoliation via thermal release tape (TRT) and subsequent transfer onto the intended substrate. The successful exfoliation of the BTO film is attributed to the notably more robust bonding force at the Ni/BTO interface than the BTO/graphene//STO interfaces. Table 1 summarises the BTO growth conditions in PLD and their impact on the feasibility of exfoliation. Notably, BTO/ML-graphene//STO failed to achieve complete exfoliation under all conditions (refer to Table 1 and Figure S 10(a) for optical images). This outcome aligns with prior research by Yuwei Guo et.al [34], they reported less than 5% yield in exfoliation for $VO_2$/monolayer graphene//sapphire stacks. Conversely, the BTO/BL-graphene//STO sample, grown via a two-step slit-based BTO deposition method, demonstrated successful exfoliation. In Figure 4(e), the specular XRD θ-2θ scan of the post-exfoliated BTO//Ni/TRT structure shows the presence of BTO (001) and (002) reflections. This substantiates the retention of the as-grown BTO orientation post-exfoliation. The existence of BTO is further corroborated by EDS signals, with discernible peaks corresponding to Ba, Ti, and O (Figure S10 (b)). However, the symmetric RC FWHM of (002) BTO increased from 0.77°±0.03° (for BTO/BL graphene//STO) to 1.1° after release, attributable to the microscopically rough surface of TRT as opposed to the STO substrate (Figure 2(b)).

**Table 1: Growth condition in PLD and feasibility of BTO exfoliation.**

| Sample | Carrier gas/Laser slit | Feasibility of exfoliation | Growth Temperature (Substrate) |
|---|---|---|---|
| BTO/BL-graphene//STO | Vacuum or $N_2$ or Ar (No slit) | N | 750 °C (STO) |
| BTO/BL-graphene//STO | Vacuum (Slit based two-step) | Y | |
| BTO/ML-graphene//STO | Vacuum or $N_2$ or Ar (Slit based two-step) | N | |

Figure 4(f) shows a post-exfoliation SEM image of the BTO surface, showcasing a planar morphology. Please take note of the rough spalling marks observed in small regions, which stem from direct epitaxy due to localized defects or holes in the graphene. This issue needs to be resolved by enhancing the efficiency of graphene transfer. After exfoliation, Raman spectroscopy was conducted on the STO substrate and the exfoliated BTO surface to ascertain the graphene layer's location, as detailed in Figure 4(g, k). The Raman spectrum of the STO substrate exhibits peaks corresponding to the D, G, and 2D peaks of the graphene layer, indicating the graphene's presence on STO post-exfoliation (Figure 4(k)). Conversely, the Raman spectrum of the exfoliated BTO layer exhibits no characteristic graphene peaks within the 1200-3000 $cm^{-1}$ range, as depicted in Figure 4(g). Figure 4(i, j) presents the post-exfoliation STO substrate optical and SEM images. A clear demarcation, segregating the graphene region from the non-graphene region (which lacks top film exfoliation), is observable (see Figure 4(j) SEM image, also refer to Figure S11 and Figure S12 for optical images). A graphene//Cu, measuring 4mm x 5mm, was sectioned from a 1inch x3 inches graphene//Cu foil sheet and transferred onto a 5mm x 5mm STO substrate, effectively delineating the failed exfoliation zones without graphene.

Piezoelectric force microscopy (PFM) was employed to investigate the ferroelectricity of the exfoliated membrane on TRT, with Ni serving as the underlying electrode. A bias of +7 V was applied over an outer region of 3.5 μm x 3.5 μm, followed by a subsequent -7 V application within a central 1.5 μm x 1.5 μm region. Subsequently, a small AC voltage of 0.7 V was applied to the tip to discern variations in polarity. The results are illustrated in Figure 4(h), depicting PFM phase and amplitude maps. The phase map distinctly displays a 180° contrast between oppositely poled regions, confirming ferroelectricity in the FS BTO film.

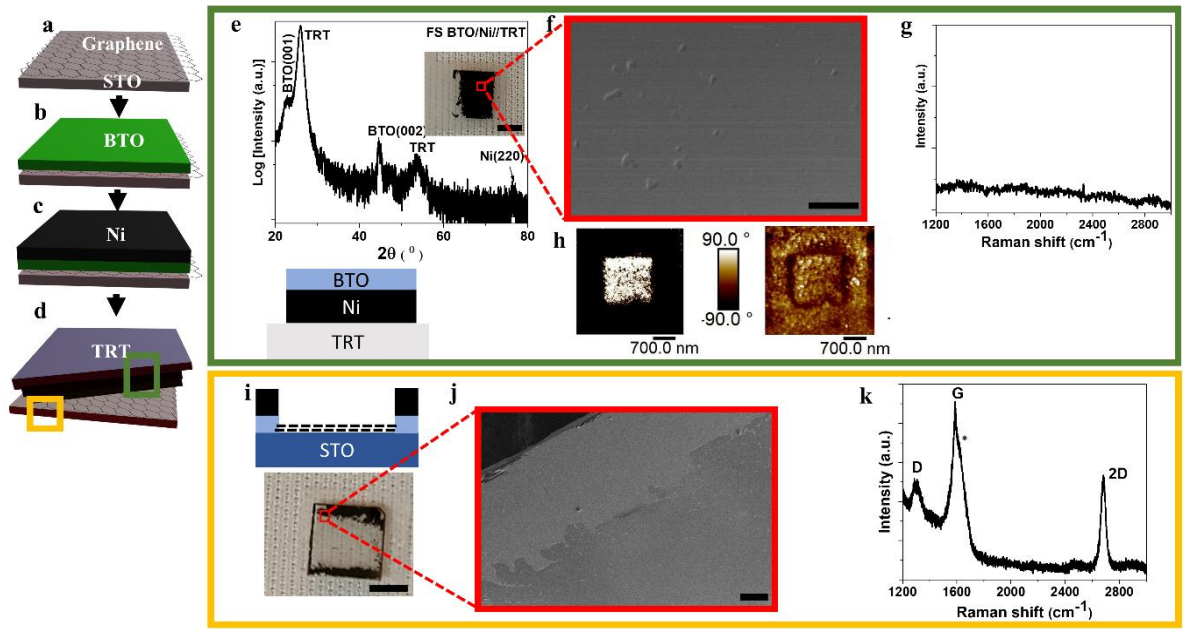

**Figure 4: Exfoliation of BTO membrane.** (a-d) Schematic showing the complete process starting from PLD BTO on graphene//STO, followed by micron thick Ni stressor layer deposition and exfoliation using thermal release tape (TRT); (e) XRD θ-2θ between 20-80° of BTO post exfoliation (BTO/Ni//TRT stack), inset showing an optical image of the stack (scale bar 2.5 mm and also see Figure S12 for optical images), (f) SEM image of the BTO surface post exfoliation (scale bar 25 μm), (g) Raman spectroscopy of BTO/Ni//TRT confirms the absence of graphene on the BTO surface, (h) PFM phase and amplitude of BTO on Ni//TRT, outer 3.5 μm x 3.5 μm area biased with 7 V and inner 1.5 μm x 1.5 μm area biased with -7 V and then a small AC voltage of 0.7 V is applied on the tip to read the contrast between the differently poled region. (i) optical image of STO substrate post exfoliation of BTO (scale bar 2.5 mm). Notably, regions without graphene (located near the two edges) exhibit a lack of successful BTO exfoliation, as demonstrated in the magnified image in (j) SEM (scale bar of 200 μm), (k) Raman spectroscopy of STO substrate post exfoliation of BTO showing characteristics peak of graphene D G and 2D peaks.

Subsequently, the BTO/Ni//TRT stack was transferred onto a SiO$_2$ substrate by stamping it, and TRT was eliminated by heating the substrate to 110 °C. The cross-sectional HAADF-STEM image of the Ni/BTO//SiO$_2$ stack is shown in Figure 5(a). The HRTEM image of FS BTO with Fast Fourier Transformations (FFT) taken from a selected region confirms the transfer of crystalline FS BTO onto the foreign substrate (SiO$_2$) (Figure 5(b, c)). The STEM-EDS composition mapping of the transferred film onto the SiO$_2$ substrate is visualized in the provided Figure 5(d-h). However, a limited fraction of film transferred from TRT to SiO$_2$, can be improved by using eutectic or surface activation-based wafer bonding techniques.

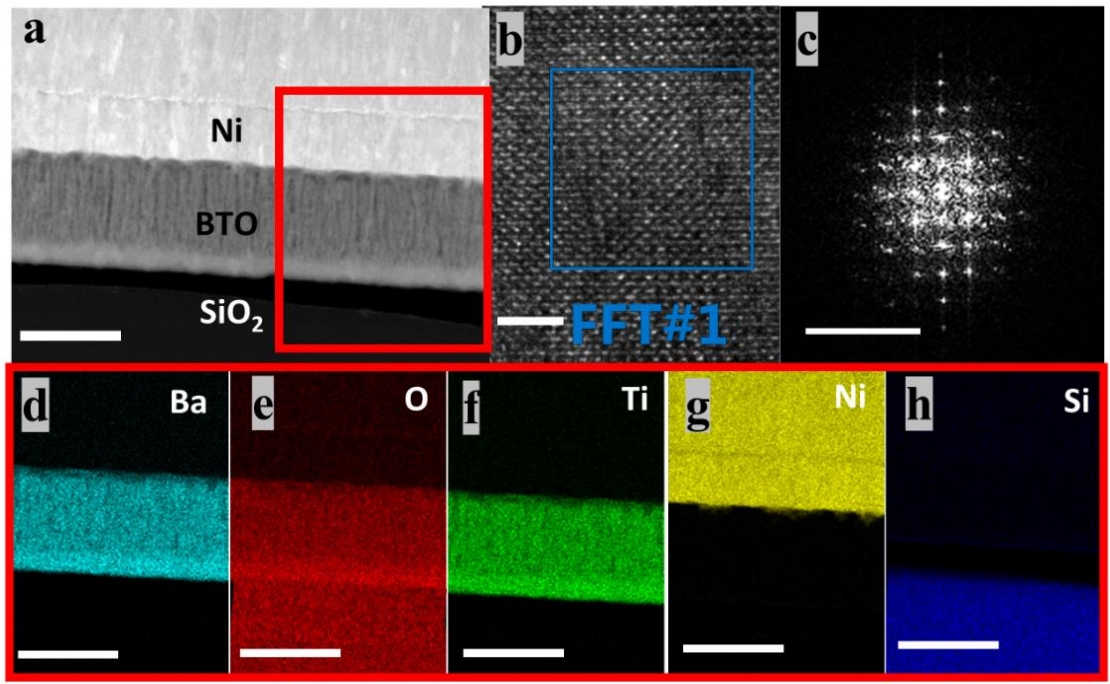

**Figure 5: TEM characterization of transferred BTO on SiO$_2$:** (a) cross-sectional HAADF-STEM image (scale bar 100 nm), (b, c) HRTEM (scale bar 2 nm) and FFT in the BTO region (scale bar $10^{-1}$ nm), (d-h) STEM-EDS mapping of the selected area on (a) (red box), scale bar 100 nm.

To understand the underlying mechanism of exfoliation, it is significant to explore the strain relaxation of BTO with the number of layers of graphene at the interface. The out-of-plane (OP) and in-plane (IP) lattice parameters were derived from XRD θ-2θ BTO (002) and BTO (103) diffraction peaks, respectively, as shown in Figure S9 for BTO/(ML, BL, TL) graphene//STO configurations. Figure 6(a) depicts the lattice parameter's evolution, illustrating its dependence on the number of graphene layers. Lattice parameters of the bulk freestanding BTO film are used as the internal reference here. It is to be noted that the BTO being defective, as earlier reported [6], the internal reference lattice parameters are different from that of bulk defect free BTO. It can be seen that the BTO film on STO is strained in comparison to the freestanding film. Introducing graphene layers at the interfaces reduces this strain and makes it almost comparable to that of the freestanding film when two layers are used. The introduction of the graphene layer allowed atomic registry guidance due to remote interactions, and the attenuated binding energy between the epilayer and substrate facilitated the relaxation of the misfit strain in the epitaxial film grown on graphene [27]. The resultant outcome of the reduced interaction between the BTO and STO due to the bi-layer graphene (as depicted schematically in the bottom row of Figure 6) helps in complete exfoliation.

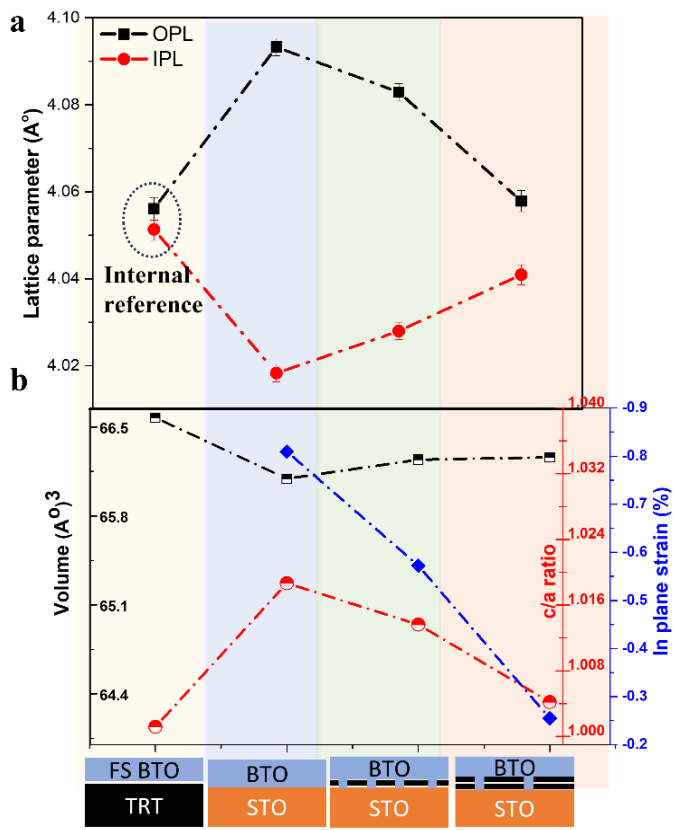

**Figure 6: Structural characterization of BTO on multilayer graphene-coated STO substrate.** (a) Evolution of lattice parameter as a function of the number of graphene layers at the interface. (OP: out-of-plane lattice parameter, IP: in-plane lattice parameter). (b) The evolution of c/a ratio (red circles), lattice volume (black squares), and in-plane strain (blue squares) in the film as a function of the number of graphene layers. In-plane strains were calculated using freestanding (FS) BTO lattice parameters as a reference. The lower schematic row illustrates the BTO on STO heterostructure, progressing from left to right: FS BTO on thermal release tape (TRT), absence of graphene, monolayer graphene, and bilayer graphene at the interface.

## Conclusion:

In conclusion, we studied the stability of wet-transferred graphene when subjected to a pulsed laser deposition (PLD) of BTO plume at a high temperature of 700°C, all while in the presence of nitrogen ($N_2$) or argon (Ar) gases at a pressure of $5 \times 10^{-3}$ mbar. Our investigation led us to propose a two-step deposition process for BTO. Initially, we employed a slit at the laser source, resulting in a growth rate of 0.02 nm/pulse, followed by a second step without the slit, yielding a growth rate of 0.08 nm/pulse. This aimed to minimize the graphene defect peak's Full Width at Half Maximum (FWHM). Our results show that the size of graphene grains, which we controlled by using different types of copper foil during the graphene growth process, significantly affects how well graphene withstands the impact of the BTO plume. This, in turn, influences the quality of the BTO film that grows on top. Our study extended to exploring the strain relaxation dynamics when incorporating multilayer graphene at the interface of the BTO/graphene//STO heterostructure. Intriguingly, epitaxy was maintained up to the bilayer of graphene, beyond which the BTO is polycrystalline. Importantly, we found that a bilayer graphene interface facilitated more extensive exfoliation of ferroelectric BTO, while monolayer graphene led to failed exfoliation. Further studies will explore the better crystalline quality of the oxide films by directly growing graphene on STO and exploring other 2D materials like $MoS_2$ and hBN for PLD oxide remote epitaxy.

## Materials and Methods:

### Graphene growth:
Graphene was synthesized on copper foils using an in-house built hot-walled low-pressure chemical vapor deposition (LPCVD) reactor with ultra-high purity hydrogen, methane, and argon as precursors. The reactor pressure was maintained at 0.97 Torr throughout the growth process. It is established that the graphene grain size increases with a higher oxygen concentration in the copper substrate [35]. Large and small grain sizes were obtained using Alfa Aesar 46365 and oxygen-free (O.F.H.C.) GF65230972 copper foils, respectively.

For the growth of large grains, the reaction chamber containing Alfa Aesar 46365 copper foil was heated from room temperature to 1040°C in an argon ambient. This process allowed a baseline oxygen influx (leak) of $1\times10^{-5}$ sccm from the environment surrounding the reaction chamber, thereby increasing the residual oxygen concentration in the copper substrate. Conversely, the reaction chamber containing oxygen-free GF65230972 copper foil was heated to 1040°C in a hydrogen ambient to maintain a reducing environment within the chamber.

After the temperature ramp-up, both copper foils were annealed in a hydrogen ambient for 60 minutes. Graphene growth was subsequently conducted using 1.05 sccm of methane and 150 sccm of hydrogen. Post-growth annealing (PGA) was performed on the full-coverage graphene with large grain sizes obtained on Alfa Aesar 46365 copper by increasing the methane flow rate to 3 sccm to heal grain boundaries [24]. During the temperature ramp-down step, the methane flow rate was maintained at 3sccm along with 150sccm of hydrogen to prevent etching of the graphene grain boundaries by hydrogen. For graphene with small grain sizes obtained on oxygen-free GF65230972 copper, the temperature ramp-down step was carried out in a hydrogen ambient. (Refer to Figure S5 for steps involved in the growth process of large and small grain graphene with Figure S6 showing the SEM image of partial coverage of graphene and method of graphene grain boundary determination.)

**Graphene Transfer:**

For a detailed step-by-step process, refer to Figure S1. The as-grown graphene on the copper foil was spin-coated with PMMA at a speed of 2000 revolutions per minute (rpm) and left to cure at room temperature for 24 hours. The graphene layer on the bottom side was then removed using a solution of 1:1 $HNO_3$ to deionized (DI) water. An ammonium persulphate solution was employed to etch the copper foil. The remaining graphene, along with the PMMA layer, was transferred into DI water to dissolve any salt residue. The cleaned graphene was subsequently collected and dried on the desired substrate. The PMMA layer was removed by immersing it in acetone overnight and then rinsing it with fresh acetone and isopropyl alcohol to eliminate any PMMA residue.

**BTO growth for remote heteroepitaxy:**

The BTO target was created by pelletizing and sintering nano-powders of BTO at 1350 °C for 12 hours. The BTO thin film was grown on the with/without graphene-coated STO substrate

(5mm x 5mm substrates of STO procured from Shinkosha Co.Ltd., Japan) inside a reflection high-energy electron diffraction (RHEED)-assisted pulsed laser deposition (PLD) chamber, utilizing a KrF excimer laser with a wavelength of 248 nm. The BTO target was ablated with a laser fluence of 1.04 J/cm$^2$, a laser frequency of 1 Hz, and a substrate temperature of 700-750 °C following two-step deposition, as depicted in Figure 1(d). A constant substrate-to-target distance of 5 cm was maintained for all deposition processes.

**BTO exfoliation using Ni Stressor:**

To transfer BTO, a Ni stressor layer was first grown by initial 100 nm by an e-beam evaporator and subsequently by sputtering 1 μm thick Ni layer. Thereafter, with the help of TRT BTO/Ni layer was separated from the STO substrate. Figure 4(a) provides a schematic representation of the transfer process.

**Characterization:**

X-ray diffraction (XRD) analysis was performed using a four-circle XRD instrument (Rigaku smart lab) equipped with a CuKα source. The surface morphology ferroelectricity of the film was examined using atomic force microscopy (AFM) and Piezo force microscopy (PFM) module on a Bruker Icon Dimension system. Raman spectroscopic measurements were conducted using a LabRAM HR spectrometer with a 532 nm laser source to investigate the presence of the graphene layer. Cross-sectional transmission electron microscopy (TEM) was employed to analyze a lamella prepared using focused ion beam (FIB) techniques. The lamella was then examined using TEM (Titan G2) at an operating voltage of 300 kV. The chemical analysis of the film was conducted via X-ray photoelectron spectroscopy (XPS), employing a monochromatic aluminium source within a Kratos Axis Ultra XPS system.

**AUTHOR INFORMATION**

**Corresponding Authors**

*Srinivasan Raghavan, Email: sraghavan@iisc.ac.in

*Asraful Haque, Email: asrafulhaque@iisc.ac.in

**Present Addresses**

*Center for Nanoscience and Engineering, Indian Institute of Science, Bengaluru, India, 560012

**Notes**

The authors declare no competing financial interest.

**ASSOCIATED CONTENT**

**Data availability**: The data supporting this article have been included as part of the **Supplementary Information**.

**ACKNOWLEDGMENT**

This work was partly carried out at the Micro and Nano Characterization Facility (MNCF) and National Nanofabrication Center (NNfC) located at CeNSE, IISc Bengaluru, funded under Grant DST/NM/NNetRA/2018(G)-IISc, NIEIN, and Ministry of Human Resource and Development, Government of India and benefitted from all the help and support from the staff. AH would like to thank Dr. Sandeep Vura for the initial discussions.

# Supplementary Information

# Free Standing Epitaxial Oxides Through Remote Epitaxy: The Role of the Evolving Graphene Microstructure

Asraful Haque*, Suman Kumar Mandal, Shubham Kumar Parate, Harshal Jason D'souza, Sakshi Chandola, Pavan Nukala, Srinivasan Raghavan*.

Center for Nanoscience and Engineering, Indian Institute of Science, Bengaluru.

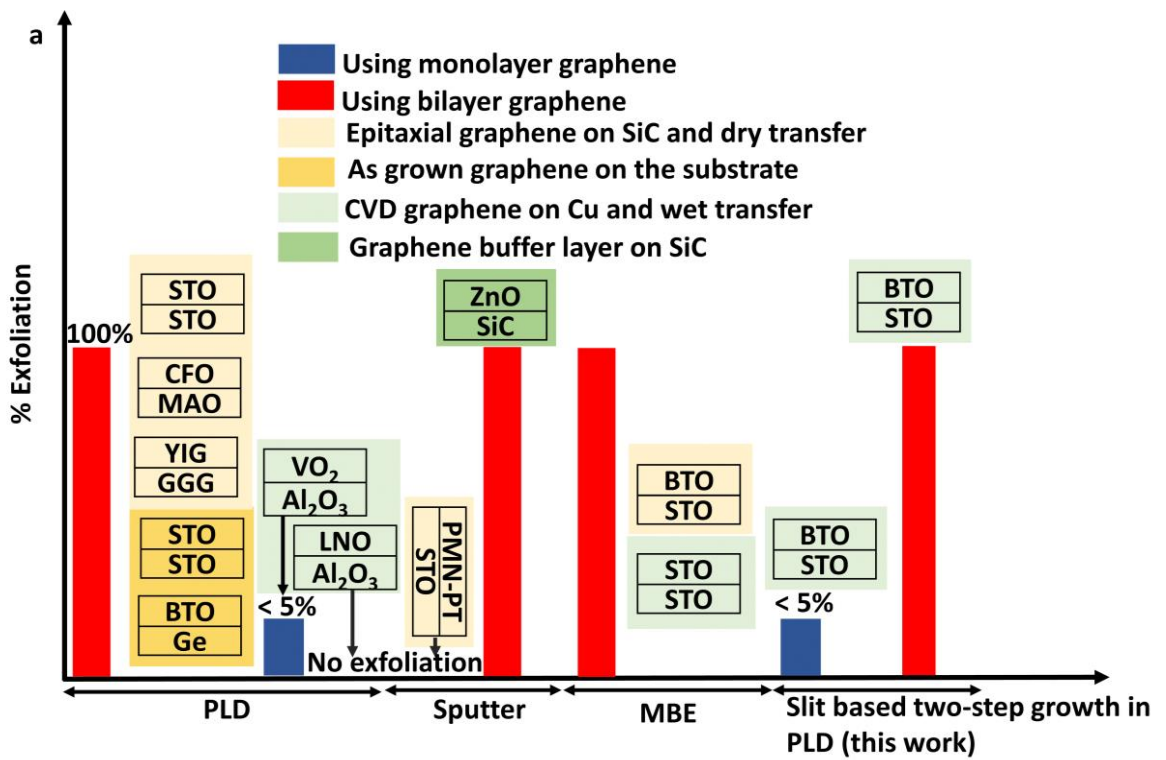

**Figure S1**: **The epitaxial exfoliation of functional oxides utilizing diverse growth methodologies such as PLD, sputtering, and MBE, facilitated through remote epitaxy by employing graphene as reported in the literature** [1–6]. Noteworthy is the utilization of graphene, either transferred via wet or dry techniques or directly grown onto the substrate of interest, as distinguished by distinct colour codes. Instances,

wherein a bilayer of graphene was employed are delineated in red and blue for monolayer graphene. (CoFe$_2$O$_5$, CFO; MgAl$_2$O$_4$, MAO; Y$_3$Fe$_5$O$_{12}$, YIG; Gd$_3$Ga$_5$O$_{12}$, GGG; LiNbO$_3$, LNO; Pb(Mg$_{0.33}$Nb$_{0.67}$)O$_3$-PbTiO$_3$, PMN-PT. )

**Section I: Transfer of CVD graphene on copper foil to other substrates of interest:**

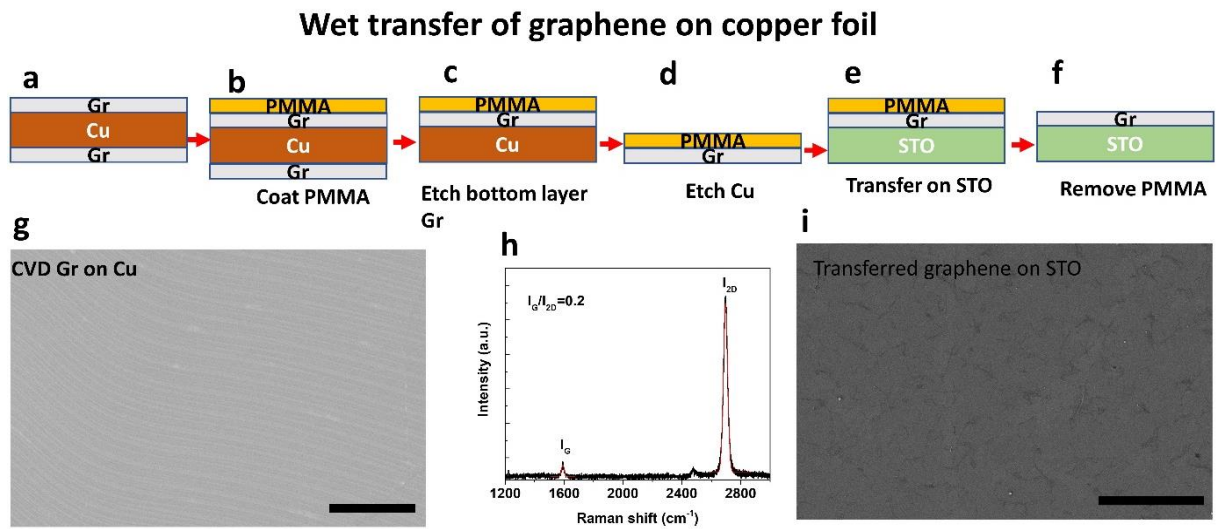

**Figure S2: Wet transfer of CVD graphene** (a) CVD grown graphene on copper substrate, (b) Spin coating of PMMA at 2000 rpm for 60 sec, (C) Backside etching of graphene in a 1:1 HNO$_3$:DI water solution for 10 seconds, (d) copper etching in ammonium persulphate solution, after that PMMA/graphene layer was transferred onto fresh DI water and kept there for 10 minutes to dissolve salt residues, (e) PMMA/graphene was scooped with a 5mm x 5mm STO or SiO$_2$/Si substrate depending on the usage, (f) PMMA was removed by dipping the sample inside acetone for overnight followed by cleaning with IPA solution, (g) SEM image of CVD grown graphene on copper foil. Copper step beams are observed, which arises due to the release of thermal stress between grown graphene and copper foil [7], (h) Raman spectra between 1200-3000 cm$^{-1}$ of graphene on SiO$_2$/Si substrate indicating graphene G and 2D peaks with monolayer characteristics, (i) SEM image of as transferred graphene on STO showing PMMA residue. The scale bar in Figures (g) and (i) is 50 µm.

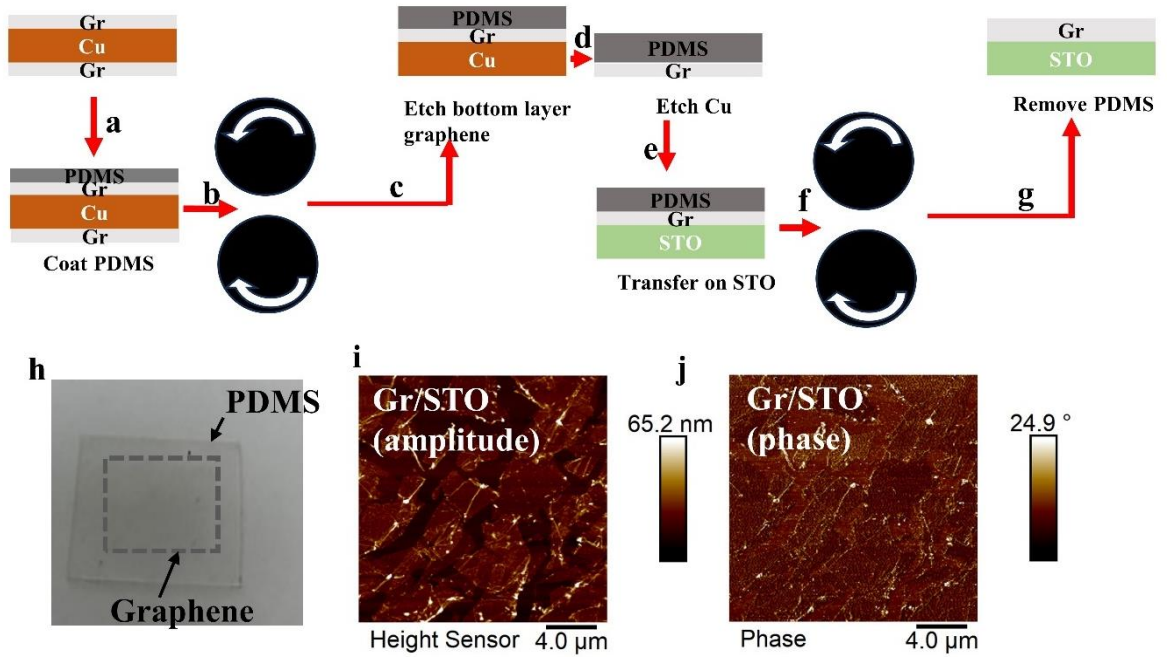

**Figure S3: Dry transfer of graphene:** On CVD-grown graphene on copper foil (a-c) PDMS sheet (procured from GelPak) was stamped using roll bonding technique at room temperature, and backside graphene was etched using HNO3: DI water (1:1) solution, (d) copper etching using ammonium persulphate solution, (e,f) PDMS with graphene was roll bonded to STO at room temperature followed by peeling of PDMS. (h) Optical image of graphene on PDMS. AFM (i) amplitude (j) phase micrograph of graphene on STO obtained through the dry transfer process. Graphene tears during the transfer process due to the application of mechanical force.

**Raman Characterization of Graphene**

**Table TS1: Effect of gases and slit on FWHM of the D peak of graphene and $I_D/I_G$ ratio.**

| BTO/Gr//SiO$_2$ at 700 $^0$C (mbar) | FWHM of D peak (in cm$^{-1}$) ($I_D/I_G$) | | |
|---|---|---|---|
| | Point 1 | Point 2 | Point 3 |
| with slit (5x10$^{-6)}$ | 24.4 ± 0.5 (1.2) | 23.1 ± 0.4 (1.3) | 25.1 ± 0.3 (1.3) |
| without slit (5x10$^{-6}$) | 72.6 ± 0.8 (1.1) | 65.9 ± 1.4 (1.2) | 66.3 ± 1.1 (1.2) |
| N₂ (5x10⁻³) | 79.6 ± 1.2 (1.7) | 87.7 ± 2.7 (1.6) | 89.7 ± 1.3 (1.4) |
| Ar (5x10⁻³) | 67.4 ± 1.2 (1.2) | 69.6 ± 2.3 (1.1) | 61.3 ± 1.5 (1.2) |

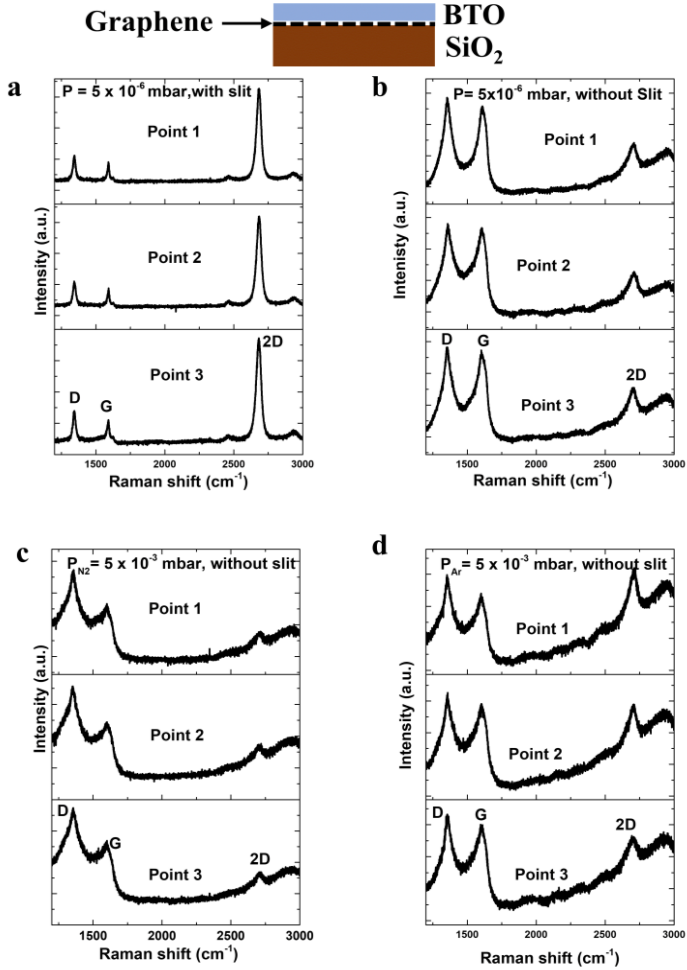

**Figure S4: Raman spectra of graphene on SiO₂/Si post-treatment inside PLD chamber:** at temperature of 700 °C and pressure of (a) 5x10⁻⁶ mbar, with slit at the laser, (b) 5x10⁻⁶ mbar, without slit at the laser, (c) 5x10⁻³ in N₂ ambient, without a slit at the laser, (d) 5x10⁻³ mabr Ar ambient, without a slit at the laser, and their corresponding FWHM of the D peak in Table 1 (row 1 to 4, respectively).

**Section II: Graphene growth:**

The expression for supersaturation (ΔG) of the graphene formation reaction is given in equation 1, where $\eta_{CH_4}$, $\eta_{H_2}$ and $\eta_{total}$ correspond to the methane, hydrogen, and the total flow rate of gases, respectively. $P_{total}$ is the reactor pressure, T is the reactor temperature, and $K_{eq}$ is the equilibrium constant for the graphene formation reaction.

$$CH_4 \leftrightarrow C_{Graphene} + 2H_2 \qquad (1a)$$

$$\Delta G = RT \ln\left(\frac{\eta_{CH_4} * \eta_{total}}{\eta_{H_2}^2 * P_{total} * K_{eq}}\right) \qquad (1b)$$

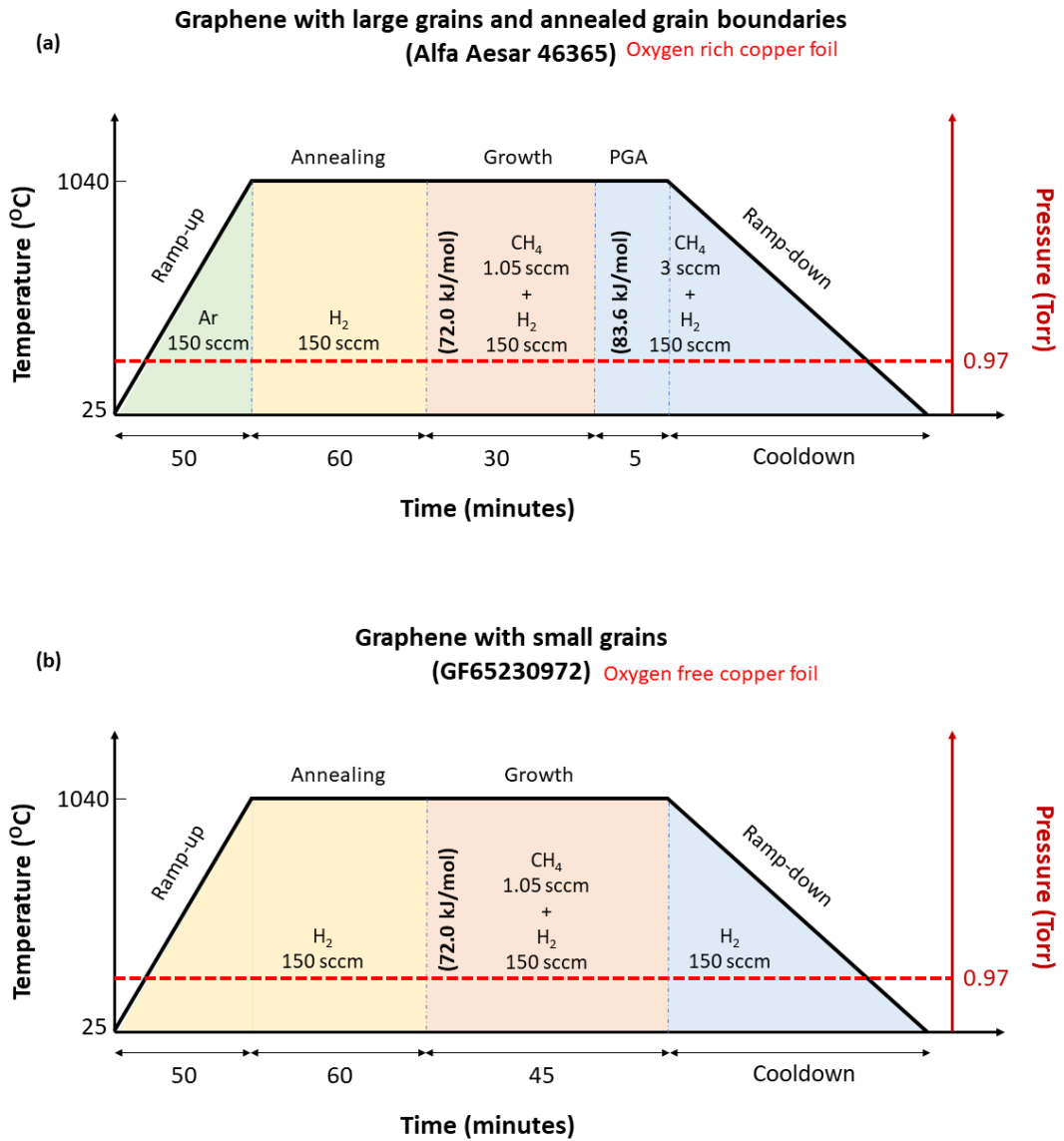

**Figure S5: Graphene growth** for (a) large grain and (b) small grain sizes are presented schematically here. The gas flow is mentioned in SCCM during ramp-up, annealing, and post-growth annealing (PGA). Different supersaturations are maintained during the growth of large and small grain graphene as mentioned in kJ/mol and calculated using Equation (1).

**Determination of the grain size:**

The average grain size of the graphene crystals can be calculated by measuring the saturation nucleation density ($N_{sat2D}$) of the crystals on the copper surface. No more crystals can form on the substrate once $N_{sat2D}$ is reached, and thus, only the existing crystals grow and merge to form a full-coverage film with time [8]. The area per grain in the final full-coverage film is then the reciprocal of $N_{sat2D}$. Assuming that the final merged crystals have an approximate circular disk shape, the average diameter of the graphene grains is then calculated (Equation 2)[8].

$$d_{grain} = \frac{2}{\sqrt{\pi\, N_{sat2D}}} \qquad (2)$$

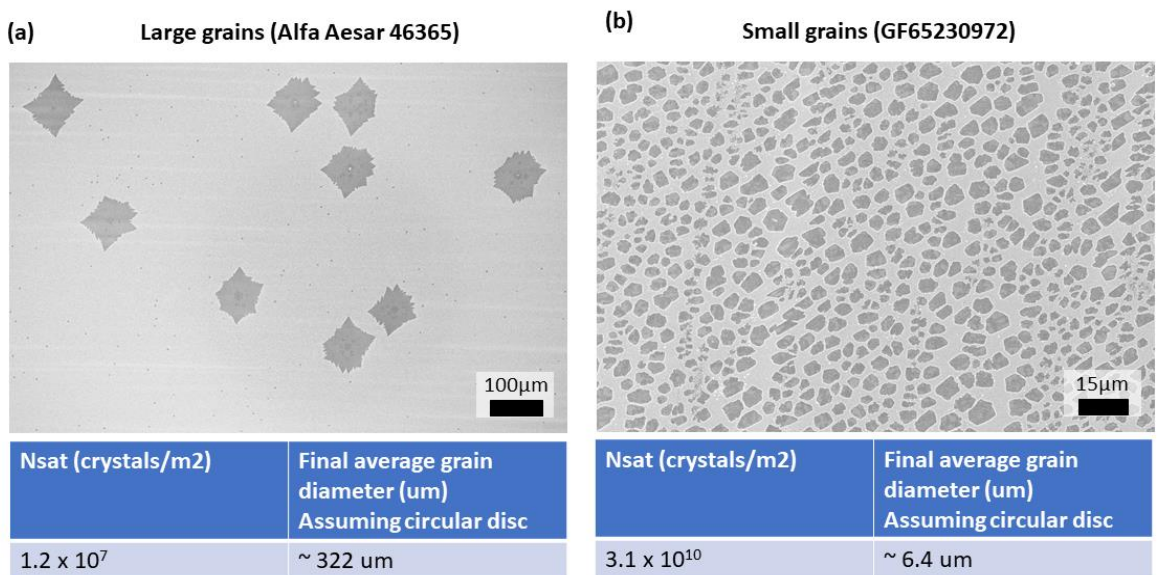

Figure S6: **SEM image of graphene with partial coverage** of (a) large grain following growth steps as in Figure S5(a) and (b) small grain following growth steps as in Figure S5(b).

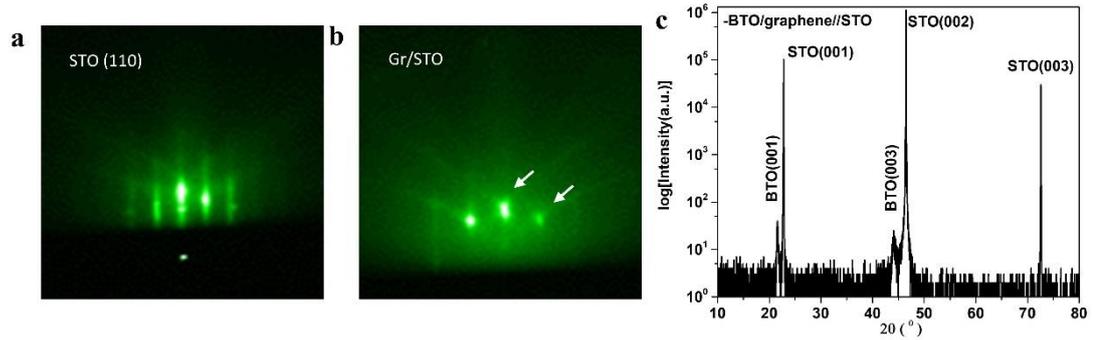

**Figure S7: RHHED pattern** of (a) graphene-coated STO substrate and (b) bare STO substrate. The white arrows in (a) indicated the spots from underneath the STO substrate. (c) HRXRD θ-2θ scan between 10-80⁰ for BTO/graphene/STO.

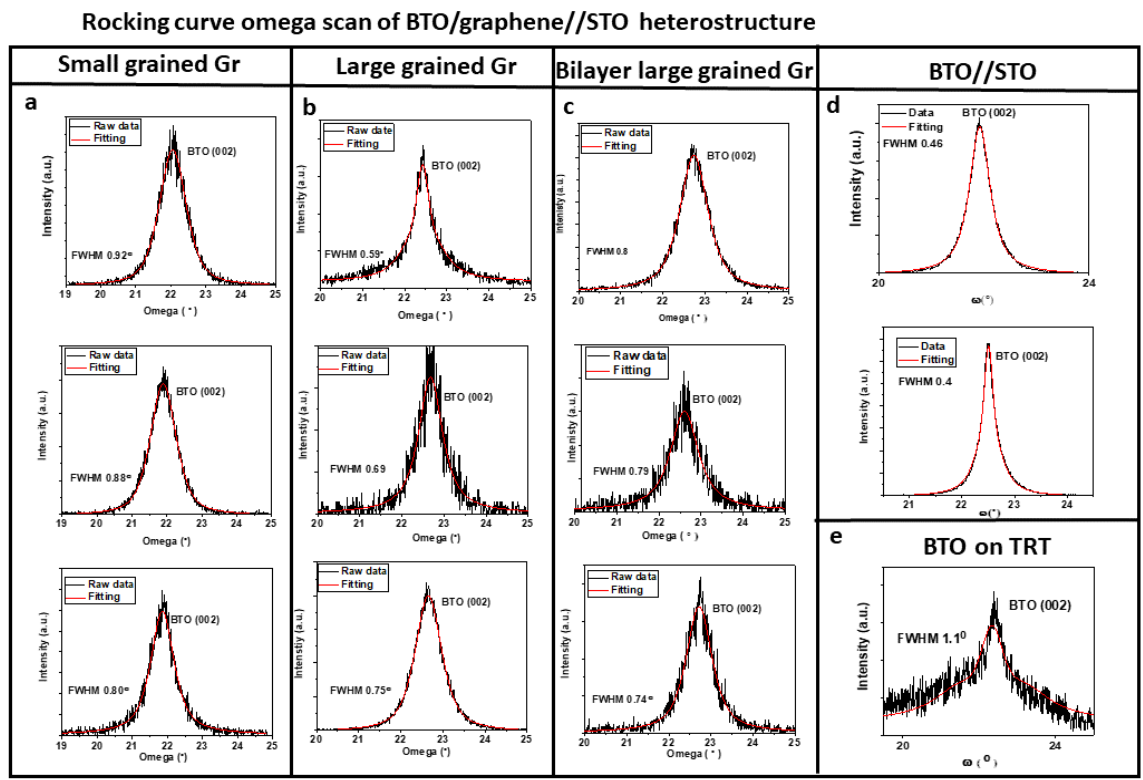

**Figure S8: Rocking curve FWHM of BTO (002) grown on** (a) small grain graphene (s-gr)~ 6.4 µm//STO, (b) large grain graphene (L-gr)~ 322µm //STO, (c) bilayer large grained graphene (BL L-gr)//STO, (d) on STO and (e) freestanding (FS) BTO on TRT.

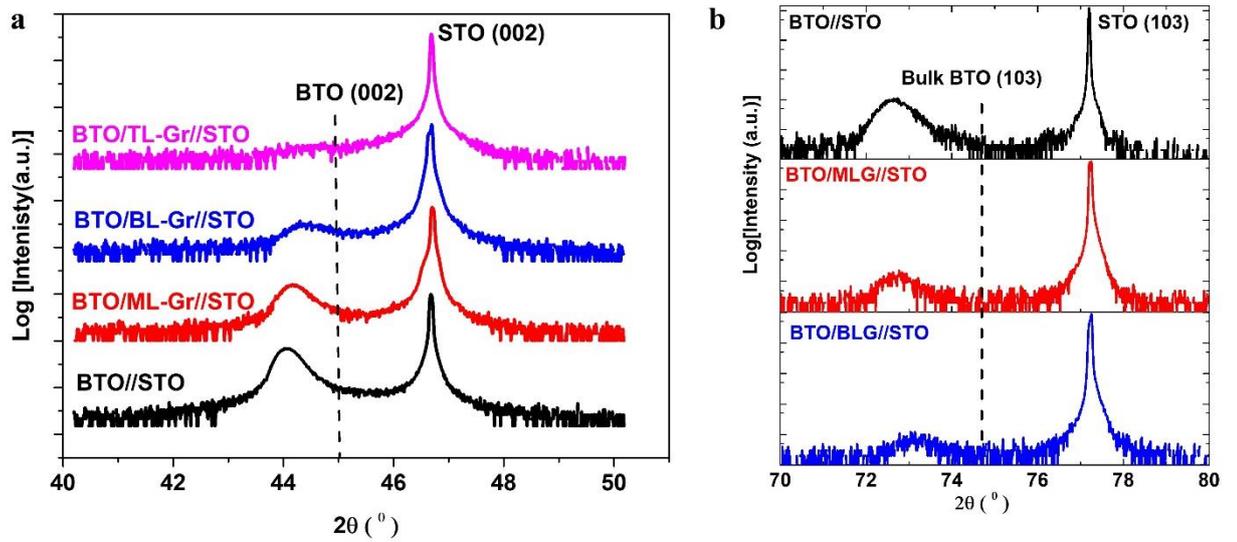

**Figure S9:** (a) Specular out-of-plane (OP) XRD θ-2θ scan of BTO/(ML, BL, TL) graphene//STO and BTO//STO in the proximity of STO (002) Bragg reflection, (b) around asymmetric (103) Bragg reflection. Here, ML-gr monolayer graphene, BL-gr bilayer graphene, TL-gr trilayer graphene.

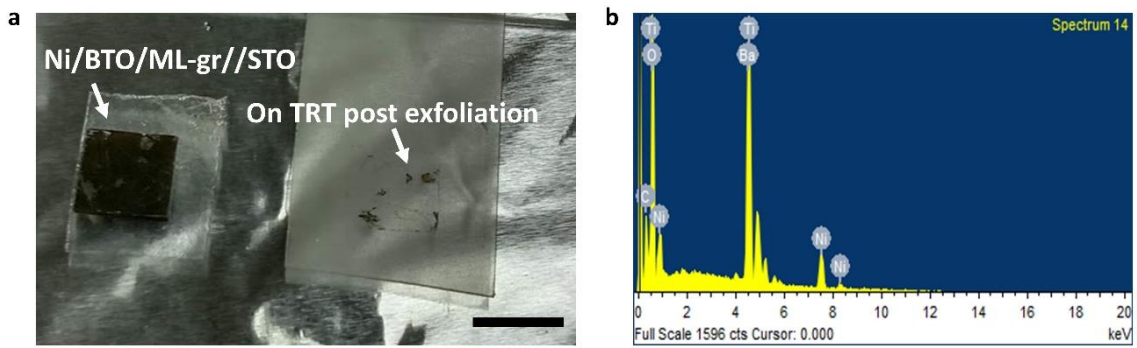

**Figure S10: Exfoliated membrane:** (a) optical image of partial exfoliation of BTO from BTO/monolayer graphene//STO stack. (b) SEM-EDX from the exfoliated BTO from BTO/bilayer graphene//STO stack.

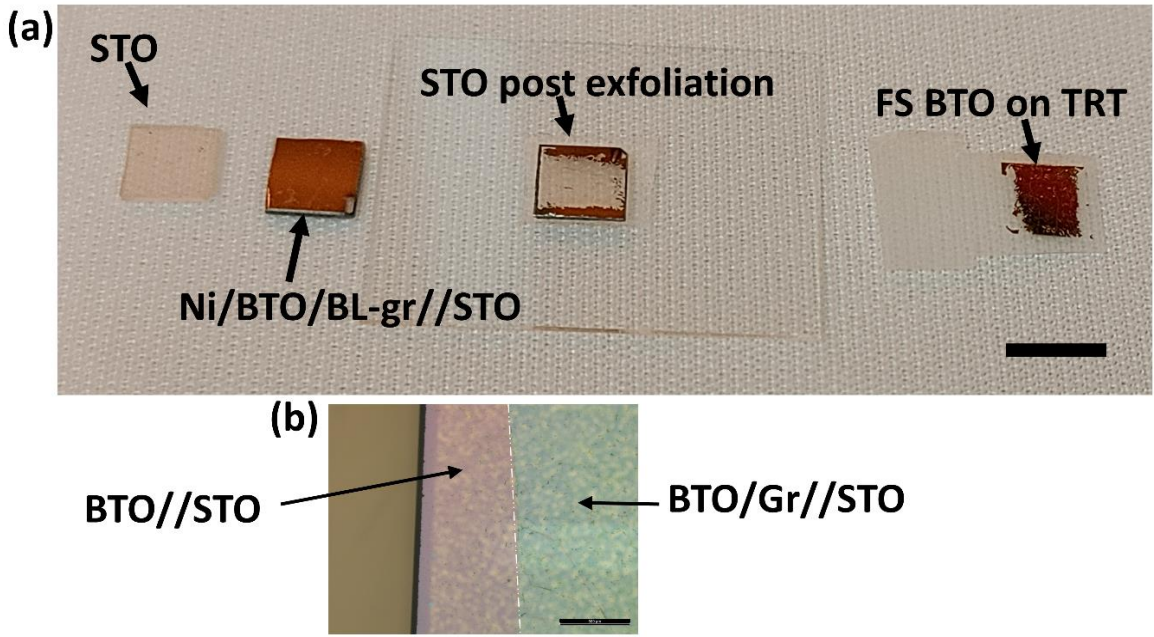

Figure S11: **(a) Optical image showing the exfoliation process of BTO using large-grained bilayer graphene** (scale bar 5mm). (b) An optical image of BTO/Gr//STO shows the region with and without graphene near the sample edge.

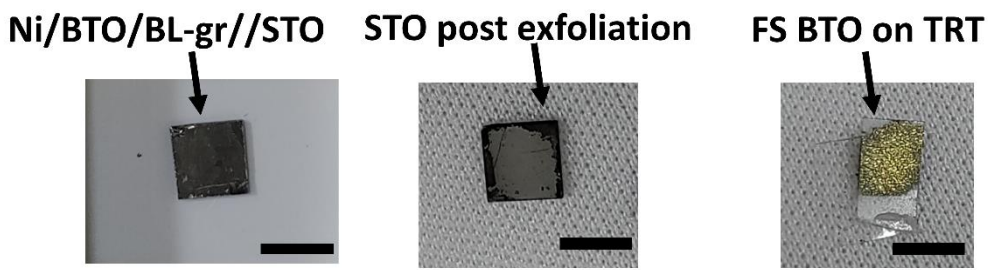

Figure S12: **Optical image showing the exfoliation process of BTO using small-grained graphene** (scale bar 5 mm).

**Strain relaxation with graphene layers at the interface:**

Compared to BTO//STO, BTO/ML graphene//STO resulted in a slight increase in the IP lattice parameter to 4.03 A° (Figure 6(a)). This corresponded to a 0.24% reduction in in-plane strain, while the OP lattice parameter decreased (see Figure 6(b)). Strain calculations were based on the lattice parameters of exfoliated BTO, serving as a reference for bulk defective BTO, with IP and OP values of 4.05 A° and 4.06 A°, respectively (Figure 6(a)). With the addition of a second graphene layer (BTO/BL-graphene//STO), the in-plane strain was reduced by 0.32%, resulting in OP and IP lattice parameters of 4.057 A° and 4.04 A°, respectively (as shown in Figure 6(a, b)).

The reduction of in-plane strain was explained by the interaction between the graphene layers, enabling relaxation. The OP and IP lattice parameters decreased to 4.06 A° and 4.05 A° upon film exfoliation, respectively (Figure 6(a)).

Along with a gradual increase in in-plane strain relaxation with the number of graphene layers, as shown in Figure 6(b), leads us to the conclusion that the number of direct epitaxial interactions of BTO through defect sites in graphene is diminished in the presence of multiple layers of graphene compared to a monolayer at the interface (refer to the schematic in the bottom row of Figure 6(b)), thus aiding the exfoliation process in the former case. A comprehensive evaluation of lattice volume and c/a ratio was conducted as a function of the number of graphene layers (Figure 6(b)). Compared to BTO//STO with a c/a ratio of 1.02, the ratio decreased to 1.01 and 1.004 upon introducing ML and BL graphene, respectively. A further reduction to 1.001 was observed after film exfoliation. Additionally, there is a minimal lattice volume expansion from 66.09 (A°)$^3$ (BTO//STO) to 66.24 (A°)$^3$ (BTO/ML graphene//STO) and 66.26 (A°)$^3$ (BTO/BL graphene//STO), culminating in 66.57 (A°)$^3$ after exfoliation.